\documentclass[twocolumn,showpacs,preprintnumbers,amsmath,amssymb,superscriptaddress]{revtex4}

\usepackage{graphicx}
\usepackage{dcolumn}
\usepackage{bm}

\begin{document}

%

\def\theend{rip}
\def\TheEnd{Big Rip}

%

\title{Phantom Energy and Cosmic Doomsday}

\author{Robert R. Caldwell}
\affiliation{Department of Physics \& Astronomy, Dartmouth College,
6127 Wilder Laboratory,
Hanover, NH 03755}

\author{Marc Kamionkowski}
\affiliation{Mail Code 130-33, California Institute of Technology,
Pasadena, CA 91125}

\author{Nevin N. Weinberg}
\affiliation{Mail Code 130-33, California Institute of Technology,
Pasadena, CA 91125}


\begin{abstract}
Cosmologists have long wondered whether the Universe will eventually
re-collapse and end with a Big Crunch, or expand forever, becoming increasingly
cold and empty.  Recent evidence for a flat Universe, possibly with a
cosmological constant or some other sort of negative-pressure dark energy, has
suggested that our fate is the latter.  However, the data may actually be
pointing toward an astonishingly different cosmic end game.  Here, we explore
the consequences that follow if the dark energy is phantom energy, in which the
sum of the pressure and energy density is negative. The positive phantom-energy
density becomes infinite in finite time, overcoming all other forms of matter,
such that the gravitational repulsion rapidly brings our brief epoch of cosmic
structure to a close. The phantom energy rips apart the Milky Way, solar
system, Earth, and ultimately the molecules, atoms, nuclei, and nucleons of
which we are composed, before the death of the Universe in a ``\TheEnd''. 
\end{abstract}

\maketitle
\smallskip 

Hubble's discovery of the cosmological expansion, crossed with the mathematical
predictions of Friedmann and others within Einstein's general theory of
relativity, has long sparked speculation on the ultimate fate of the Universe.
In particular, it has been shown that if the matter that fills the Universe can
be treated as a pressureless fluid, which would be the case for galaxies, then
the Universe expands forever (if it has a Euclidean or hyperbolic spatial
geometry) or eventually re-collapses (if its spatial geometry is that of a
3-sphere). Evidence from supernova searches \cite{supernovaone,supernovatwo}
and the stunning cosmic microwave background (CMB) results from balloon and
ground experiments \cite{toco,boom,max,dasi,cbi,archeops} and now from WMAP
\cite{wmapparams,wmappeaks} that indicate an accelerating cosmological
expansion show that this simple picture is not enough; the Universe
additionally consists of some sort of negative-pressure dark energy.

The dark energy is usually described by an ``equation-of-state'' parameter
$w\equiv p/\rho$, the ratio of the spatially-homogeneous dark-energy pressure
$p$ to its energy density $\rho$.  A value $w<-1/3$ is required for cosmic
acceleration. The simplest explanation for dark energy is a cosmological
constant, for which $w=-1$.  However, this cosmological constant is 120 orders
of magnitude smaller than expected from quantum gravity.  Thus, although we can
add this term to Einstein's equation, it is really only a placeholder until a
better understanding of this negative pressure arises. Another widely explored
possibility is quintessence \cite{quint,ratra,wetterich,coble,turnerwhite,spint}, a
cosmic scalar field that is displaced from, but slowly rolling to, the minimum
of its potential.  In such models, the equation-of-state parameter is
$-1<w<-1/3$, and the dark-energy density decreases with scale factor $a(t)$ as
$\rho_Q \propto a^{-3(1+w)}$.

\begin{figure}[t]
\scalebox{0.4}{\includegraphics{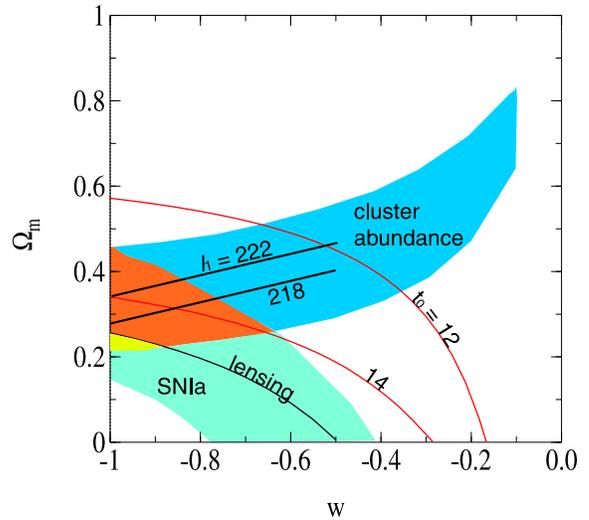}}
\caption{\label{fig:qcdm} 
     Current constraints to the $w$-$\Omega_m$ parameter
     space. The red solid curves show the age (in Gyr) of the
     Universe today (assuming a Hubble parameter $H_0=$70
     km~sec$^{-1}$~Mpc$^{-1}$).  The light shaded regions are
     those allowed (at $2\sigma$) by the observed cluster
     abundance and by current supernova measurements of the
     expansion history.  The dark orange shaded region shows the
     intersection of the cluster-abundance and supernova curves,
     additionally restricted (at $2\sigma$) by the location of
     the first acoustic peak in the cosmic-microwave-background
     power spectrum and quasar-lensing statistics.}
\end{figure}

Fig.~\ref{fig:qcdm} shows constraints to the $w$-$\Omega_m$ parameter space
(where $\Omega_m$ is the pressureless-matter density in units of the critical
density) from the cluster abundance, supernovae, quasar-lensing statistics (see
Refs. \cite{WangEtAl,perlmutter} and references therein), and the first
acoustic peak in the CMB power spectrum (values taken from Ref.
\cite{wmappeaks}). As the Figure shows, $w$ seems to be converging to $w=-1$.

But what about $w<-1$?  Might the convergence to $w=-1$ actually be indicating
that $w<-1$?  Why restrict our attention exclusively to $w\geq-1$?  Matter with
$w<-1$, dubbed ``phantom energy'' \cite{rrc}, has received increased attention
among theorists recently.  It certainly has some strange properties. For
example, the energy density of phantom energy increases with time.   It also
violates the dominant-energy condition \cite{dominant,carroll}, a cherished
notion that helps prohibit time machines and wormholes. However, it is hard to
see how time machines and wormholes would arise with phantom energy. Although
sound waves in quintessence travel at the speed of light, it does not
automatically follow that disturbances in phantom energy must propagate faster
than the speed of light; in fact, there are already several scalar-field models
for phantom energy in which the sound speed is subluminal
\cite{rrc,parker,scalarfieldsone,scalarfieldstwo,scalarfieldsthree,scalarfieldsfour}.
It is true that these models feature unusual kinetic terms in their
Lagrangians, but such terms may arise in supergravity \cite{supergravity} or
higher-derivative-gravity theories \cite{higher}.  Theorists have also
discussed stringy phantom energy \cite{frampton} and brane-world phantom energy
\cite{braneworlds}.  Connections with the dS/CFT correspondence have also been
made \cite{mcinnes}. To be sure, phantom energy is not something that any
theorist would have expected; on the other hand, not too many more theorists
anticipated a cosmological constant!  Given the limitations of our theoretical
understanding, it is certainly reasonable to ask what empirical results have to
say.

\begin{figure}[t]
\scalebox{0.4}{\includegraphics{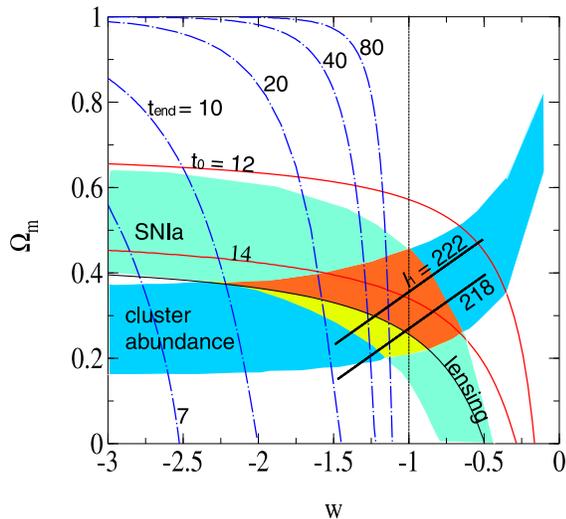}} 
\caption{\label{fig:pcdm}
     Same as in Fig.~\ref{fig:qcdm}, except extended to
     $w<-1$. Here, the blue dot-dash curves show for
     phantom-energy ($w<-1$) models the time (in Gyr) remaining
     in the Universe (assuming a Hubble parameter $H_0=70$
     km~sec$^{-1}$~Mpc$^{-1}$).}
\end{figure}

In Fig.~\ref{fig:pcdm} we generalize the analysis of cosmological constraints
to a parameter space that extends to $w<-1$.  As indicated here, there is much
acceptable parameter space in regions with $w<-1$; see also Refs.
\cite{hannestad,schuecker}.  With certain prior assumptions, the best fit is
actually at $w<-1$.

As we now show, if $w<-1$ persists, then the fate of the Universe is quite
fantastic and completely different than the possibilities previously
discussed.  To begin, let us review these other fates. In a flat or open
Universe with{\it out} dark energy, the expansion continues forever, and the
horizon grows more rapidly than the scale factor; the Universe becomes colder
and darker, but with time the co-moving volume of the observable Universe
evolves so that the number of visible galaxies  grows. If the expansion is
accelerating, as a consequence of dark energy with $-1\leq w<-1/3$, then the
expansion again continues forever.  However, in this case, the scale factor
grows more rapidly than the horizon.  As time progresses, galaxies disappear
beyond the horizon, and the Universe becomes increasingly dark.  Still,
structures that are currently gravitationally bound, such as the Milky Way and
perhaps the Local Group, remain unaffected.  Thus, although extragalactic
astronomy becomes less interesting, Galactic astronomy can continue to
thrive.\footnote{There is another possibility: if the quintessence potential at
some point becomes negative, then the Universe can reach a point of maximum
expansion and then re-collapse \cite{Steinhardt,kallosh}.}

With phantom energy, the Friedmann equation governing the time $t$ evolution of
the scale factor $a(t)$ becomes $H^2 \equiv ({\dot a}/{a})^2 = H_0^2
[\Omega_m/a^3+(1-\Omega_m) a^{-3(1+w)}]$, where $H_0$ is the Hubble parameter,
and the dot denotes a time derivative.  If $\Omega_m\simeq0.3$, then the
Universe is already dark-energy--dominated, and for $w<-1$ it will become
increasingly dark-energy--dominated in the future.  We thus approximate the
subsequent evolution of the scale factor by neglecting the first term on the
right-hand side.  Doing so, we find that the scale factor blows up in a time
$t_{\theend} -t_0 \simeq (2/3)|1+w|^{-1} H_0^{-1} (1-\Omega_m)^{-1/2}$ from the
current time $t_0$.  For example, for $w=-3/2$ and $H_0=70$
km~sec$^{-1}$~Mpc$^{-1}$, the time remaining before the Universe ends in this
``\TheEnd'' \cite{mcinnes} is 22 Gyr.

As in a cosmological-constant Universe, the scale factor grows more rapidly
than the Hubble distance $H^{-1}$ and galaxies will begin to disappear beyond
the horizon.  With phantom energy, the expansion rate $H$ grows with time, the
Hubble distance decreases, and so the disappearance of galaxies is accelerated
as the horizon closes in on us.  More intriguing is that the increase in the
dark-energy density will ultimately begin to strip apart gravitationally bound
objects.  According to general relativity, the source for the gravitational
potential is the volume integral of $\rho+3p$. So, for example, a planet in
an orbit of radius $R$ around a star of mass $M$ will become unbound roughly when
$-(4\pi/3)(\rho+ 3 p)R^{3} \simeq M$.  With $w\geq-1$, $-(\rho+3p)$ is
decreasing with time so if $-(4\pi/3)(\rho+3p) R^{3}$ is smaller than $M$
today, then it will remain so ever after.  Thus, any system that is currently
gravitationally bound (e.g., the solar system, the Milky Way, the Local Group,
galaxy clusters) will herafter remain so.

With phantom energy, $-(\rho+3p)$ increases, and so at some point in time every
gravitationally bound system will be dissociated.  With the time evolution of
the scale factor and the scaling of the phantom-energy density with time, we
find that a gravitationally-bound system of mass $M$ and radius $R$ will be
stripped at a time $t \simeq P \sqrt{2|1+3w|}/[6\pi |1+w|]$, where $P$ is the
period of a circular orbit around the system at radius $R$, before the \TheEnd~
(see Table~\ref{tab:history}).  Interestingly, this time is independent of
$H_0$ and $\Omega_m$.

\begin{table}
\caption{\label{tab:history}
The history and future of the Universe with $w=-3/2$ phantom energy.  }
\begin{ruledtabular}
\begin{tabular}{ll}
Time & Event\\
\hline  
$\sim 10^{-43}$~s  & Planck era \\ 
$\sim 10^{-36}$~s & Inflation \\
First Three Minutes & Light Elements Formed \\
$\sim 10^5$~yr & Atoms Formed \\
$\sim 1$~Gyr & First Galaxies Formed \\
$\sim 15$~Gyr & {\it Today} \\
$t_{\theend} - 1$~Gyr & Erase Galaxy Clusters\\
$t_{\theend} - 60$~Myr & Destroy Milky Way\\
$t_{\theend} - 3$~months & Unbind Solar System\\
$t_{\theend} - 30$~minutes & Earth Explodes\\
$t_{\theend} - 10^{-19}$~s & Dissociate Atoms\\
$t_{\theend} = 35$~Gyrs & \TheEnd\\
\end{tabular}
\end{ruledtabular}
\end{table}

Thus, for example, for $w=-3/2$, the interval is $t\simeq0.3\, P$ before the
end of time.  In this case, clusters will be stripped roughly a billion years
before the end of time.  In principle, if $w$ were sufficiently negative, the
Andromeda galaxy would be torn from the Local Group before it could fall into
the Milky Way; however, given current upper limits to $-w$, this is unlikely. 
For $w=-3/2$, the Milky Way will get stripped roughly 60 million years before
the \TheEnd.  Curiously, when this occurs the horizon will still be $\sim70$
Mpc, so there may still be other observable galaxies that we will also see
stripped apart (although given the time delay from distant objects, we will see
the Milky Way destroyed first).  A few months before the end of time, the Earth
will be ripped from the Sun, and $\sim 30$ minutes before the end the Earth
will fall apart.  Similar arguments also apply to objects bound by
electromagnetic or strong forces.  Thus, molecules and then atoms will be torn
apart roughly $10^{-19}$ seconds before the end, and then nuclei and nucleons
will get dissociated in the remaining interval.  In all likelihood, some new
physics (e.g., spontaneous particle production or extra-dimensional, string,
and/or quantum-gravity effects) may kick in before the ultimate singularity,
but probably after the sequence of events outlined above.

The end of structure, from cosmic, macroscopic scales down to the microscopic,
leads us to remark that our present epoch is unique from the viewpoint that at
no other time are non-linear structures possible. When the phantom energy
becomes strong enough, gravitational instability no longer works and the
Universe becomes homogeneous. Eventually, individual particles become isolated:
points separated by a distance greater than $3 {\delta t} (1+w)/(1 + 3 w)$ at a
time $t_{\theend}-\delta t$ cannot communicate before the \TheEnd. Therefore,
the dominance of the phantom energy signals the end of our brief era of cosmic
structure which began when the non-relativistic matter emerged from the
radiation.  In such a Universe, certain cosmic questions have new significance.
It is natural to find ourselves  --- or more generally,  non-linear structure
--- living close to the onset of acceleration if the structure is soon
destroyed and the Universe does not survive much longer afterwards
\cite{mcinnes}. A \TheEnd~renders the ``why now?'', or question of cosmic
coincidence, irrelevant.

The current data indicate that our Universe is poised somewhere near the
razor-thin separation between phantom energy, cosmological constant, and
quintessence. Future work, and the longer observations by WMAP, will help to
determine the nature of the dark energy.  In the meantime we are intrigued to
learn of this possible new cosmic fate that differs so remarkably from the
re-collapse or endless cooling considered before. It will be necessary to
modify the adopted slogan among cosmic futurologists --- ``{\sl Some say the
world will end in fire,\ Some say in ice}'' \cite{Frost} --- for a new fate may
await our world.

\noindent {\bf Acknowledgments}\,  
RRC thanks the UCSB KITP for hospitality.  This work was
supported at Caltech by NASA NAG5-9821 and DoE
DE-FG03-92-ER40701, at the KITP by NSF PHY99-07949, and at
Dartmouth by NSF grant PHY-0099543. NNW was supported 
by a NSF graduate fellowship.

\end{document}